\documentclass[twocolumn,showpacs,preprintnumbers,amsmath,amssymb]{revtex4}

\usepackage{graphicx}
\usepackage{dcolumn}
\usepackage{bm}

\newcommand{\bq}{\begin{equation}}
\newcommand{\eq}{\end{equation}}
\newcommand{\bqa}{\begin{eqnarray}}
\newcommand{\eqa}{\end{eqnarray}}
\newcommand{\nn}{\nonumber \\}

\def\be     {\begin{equation}}
\def\ee     {\end{equation}}
\def\bea        {\begin{eqnarray}}
\def\eea        {\end{eqnarray}}
\def\bnn    {\begin{eqnarray*}}
\def\enn    {\end{eqnarray*}}

\begin{document}

\title{Quantum phase transitions in $d-wave$ superconductors }
\author{Ki-Seok Kim, Sung-Sik Lee, Jae-Hyeon Eom and Sung-Ho Suck Salk$^{1}$}
\affiliation{Department of Physics, Pohang University of Science
and Technology Pohang 790-784, Korea \nn ${}^{1}$Korea Institute
of Advanced Studies, Seoul 130-012, Korea}
\date{\today}

\begin{abstract}
Based on an effective Lagrangian obtained from the holon-pair
boson theory of Lee and Salk [Phys. Rev. B {\bf 64}, 052501
(2001)] for high $T_c$ cuprates, we explore physical states
involved with quantum phase transitions around a critical hole
doping of $d-wave$ superconductivity. We find a new quantum phase
transition, a confinement to deconfinement transition for internal
gauge charge in the superconducting phase. An antiferro- to para-
magnetic transition in the superconducting state is explained in
the context of the confinement to deconfinement transition. We
obtain effective field theories in each region associated with the
phase transitions.
\end{abstract}

\pacs{73.43.Nq, 74.20.-z, 11.30.Rd}

\maketitle

Since experimental revelation of weak antiferromagnetism in
underdoped superconducting state\cite{Experiments}, the
coexistence of the antiferromagnetism ($AFM$) and $d-wave$
superconductivity ($dSC$) has been a subject of great theoretical
interest\cite{Theory_CD}. However, it is still unknown whether
despite the coexistence there exists any interplay (or
correlation) between the $AFM$ and $dSC$. Earlier we found that
the $AFM$ of nodal fermions can coexist with the $dSC$ with no
interplay in very low doping region\cite{Kim2_CD}. As a result the
superconducting transition in the underdoped region is found to
fall in the XY universality class in agreement with
experiments\cite{Kim2_CD,Kim1_CD,TypeII_CD}. It is naturally
expected that the $AFM$ disappears as hole concentration
increases. It is well known that in the overdoped region there
exists no $AFM$. Thus a quantum phase transition associated with
the $AFM$ is expected to occur in the superconducting phase.
Extending the earlier analysis\cite{Kim2_CD} to higher doping
where the $AFM$ is expected to disappear, we show that a
confinement to deconfinement transition ($CDT$) of internal U(1)
gauge charge can occur in the superconducting phase. This new
quantum phase transition is shown to be responsible for the
disappearance of the $AFM$ in the overdoped
region\cite{Experiments}. Further, we argue that non-Fermi liquid
behavior in high $T_c$ cuprates can be resolved in the quantum
critical point associated with the $CDT$.

Earlier, the holon-pair boson theory of Lee and
Salk\cite{LeeSalk_CD} reproduced the salient features of the
observed arch-shaped superconducting transition temperature in the
phase diagram of hole doped high $T_c$ cuprates and the
peak-dip-hump structure of optical conductivity in agreement with
observations. Following this holon-pair boson
theory\cite{LeeSalk_CD}, we introduce an effective
Lagrangian\cite{Kim2_CD} involved with low energy excitations.
Here the low energy excitations refer to the phase fluctuations of
the spinon pair and holon pair order parameters, the massless
spinons (Dirac fermions near the $d-wave$ nodal points) and holons
(bosons), and gauge fluctuations associated with these particles.
The gauge fluctuations $a_\mu$ allow the presence of internal flux
responsible for energy lowering, which arises as a result of
electron hopping. Considering the above elementary excitations, we
write the $(2+1)D$ low energy Lagrangian in the slave boson
representation, \bqa &&Z = \int
{D\psi_l}{D\psi_b}{D\phi_{sp}}{D\phi_{bp}}{Da_{\mu}}
e^{-\int{d^3x} {\cal L} } , \nn &&{\cal L} = {\cal L}_{sp} + {\cal
L}_{bp} , \nn &&{\cal L}_{sp} =
\frac{K_{sp}}{2}|\partial_{\mu}\phi_{sp} - 2a_\mu|^2 +
\bar\psi_l\gamma_\mu(\partial_\mu - ia_{\mu})\psi_l , \nn &&{\cal
L}_{bp} = \frac{K_{bp}}{2}|\partial_{\mu}\phi_{bp} - 2a_\mu -
2A_\mu|^2 + |(\partial_\mu - ia_{\mu} - iA_{\mu})\psi_b|^2 . \eqa
Here $\phi_{sp}$ ($\phi_{bp}$) is the phase field of the spinon
pair (holon pair) order parameter. $K_{sp} \sim
J_{\delta}|\Delta_{sp}|^2$ is the phase stiffness of the spinon
pair order parameter and $K_{bp} \sim
J|\Delta_{sp}|^2|\Delta_{bp}|^2$, that of the holon pair order
parameter where $J$ is the antiferromagnetic coupling and
$J_{\delta} = J(1-\delta)^2$, the renormalized antiferromagnetic
coupling in association with hole doping
$\delta$\cite{LeeSalk_CD}. $\Delta_{sp(bp)} =
|\Delta_{sp(bp)}|e^{i\phi_{sp(bp)}}$ is the spinon (holon) pairing
order parameter. $\psi_l$ represents the 4 component
spinor\cite{HerbutAF_CD,Tesanovic} of the massless Dirac fermion
near the nodal points (l = 1 and 2) and $\psi_b$, the holon
(boson) quasiparticle. $a_{\mu}$ is the U(1) internal gauge field
and $A_{\mu}$, the external electromagnetic
field\cite{Kim1_CD,TypeII_CD}.

Defining $\phi_{p} = \phi_{bp} - \phi_{sp}$ and $\phi_{c} = -
(\phi_{bp} + \phi_{sp})$\cite{Kim2_CD}, Eq. (1) is rewritten, \bqa
&&Z = \int {D\psi_l}{D\psi_b}{D\phi_{p}}{D\phi_{c}}{Da_{\mu}}
e^{-\int{d^3x} {\cal L} } , \nn &&{\cal L} = {\cal L}_{c} + {\cal
L}_{p} + {\cal L}_{int}, \nn && {\cal L}_{c} =
\frac{\kappa}{2}|\partial_\mu\phi_c + 4a_\mu + 2A_\mu|^2 +
\bar{\psi}_l\gamma_\mu (\partial_\mu - ia_\mu)\psi_l  \nn && +
|(\partial_\mu - ia_{\mu} - iA_{\mu})\psi_b|^2  , \nn &&{\cal
L}_{p} = \frac{\kappa}{2}|\partial_\mu\phi_p - 2A_\mu|^2  ,  \nn
&&{\cal L}_{int} = \kappa_{cp}(\partial_\mu\phi_c + 4a_\mu +
2A_\mu)(\partial_\mu\phi_p - 2A_\mu) , \eqa where $\kappa = \frac{
{K}_{sp} + {K}_{bp}}{4}$ and $\kappa_{cp} = \frac{ {K}_{sp} -
{K}_{bp}}{4}$. The original Lagrangian Eq. (1) is now rewritten in
terms of the Cooper pair phase field, $\phi_{p} = \phi_{bp} -
\phi_{sp}$ as a composite of the spinon pair phase field
$\phi_{sp}$ and the holon pair phase field $\phi_{bp}$, and the
"chargeon pair" phase field, $\phi_{c} = - (\phi_{bp} +
\phi_{sp})$ as a composite of the spinon pair phase field
$\phi_{sp}$ and the "anti-holon" pair phase field
$-\phi_{bp}$\cite{Kim2_CD}. Here "chargeon pair" refers to the
spin singlet pair oppositely charged to the Cooper pair and
"anti-holon", the opposite charge of holon with spin $0$. The
Cooper pair as a hole pair in the hole doped cuprates carries
charge $+2e$ and no internal gauge charge while the chargeon pair
carries charge $-2e$ and internal gauge charge $-4\tilde{e}$. Here
$\kappa$ is the phase stiffness of both the chargeon pair order
parameter $\phi_{c}$ and the Cooper pair order parameter
$\phi_{p}$.

Making a renormalization group analysis, it can be shown that via
the coupling term ${\cal L}_{int}$ the phase stiffness $\kappa$ is
renormalized in low energy limit\cite{Kim2_CD,Renormalization},
and the low energy effective Lagrangian Eq. (2) can be rewritten
as \bqa &&{\cal L} = {\cal L}_{c} + {\cal L}_{p} , \nn &&{\cal
L}_{c} = \frac{K}{2}|\partial_{\mu}\phi_c + 4a_{\mu} + 2A_{\mu}|^2
+ \bar{\psi}_l\gamma_\mu (\partial_\mu - ia_\mu)\psi_l \nn && +
|(\partial_\mu - ia_{\mu} - iA_{\mu})\psi_b|^2  , \nn &&{\cal
L}_{p} = \frac{K}{2}|\partial_{\mu}\phi_p - 2A_{\mu}|^2  , \eqa
with $K = \frac{K_{sp}K_{bp}}{K_{sp} + K_{bp}}$ where $K$ is the
renormalized phase stiffness via ${\cal L}_{int}$. The effective
Lagrangian above is now separated into two independent sectors,
one for the chargeon pair Lagrangian and the other for the Cooper
pair Lagrangian. The former explains for the internal gauge charge
$CDT$\cite{Fradkin_CD,Senthil_CD,Sudbo_CD,NaLee_CD} (as will be
discussed later) and the latter, for the superconducting phase
transition. We note that both the Dirac fermion and the holon
(boson) are coupled to the chargeon pair field via the U(1)
internal gauge field.

As hole concentration increases in the underdoped region, the
phase stiffness increases in this region\cite{Kim2_CD}. When the
hole concentration exceeds a critical value $\delta_c$ of $d-wave$
superconducting transition, the Cooper pairs are bose-condensed,
i.e., $<e^{i\phi_p}> \not= 0$ and the superconducting state
emerges. The insulator-superconductor transition at $T = 0 K$
falls into the quantum phase transition of the XY universality
class in the extreme type II limit\cite{Kim2_CD}. At finite
temperature the KT transition is expected to occur\cite{Kim1_CD}.
See Fig. 1 for a schematic diagram of showing the regions of the
Cooper pair bose condensation $<e^{i\phi_{p}}> \not= 0$ and the
chargeon pair condensation $<e^{i\phi_{c}}> \not= 0$ as a function
of hole doping $\delta$.
\begin{figure}
\includegraphics[width=6cm]{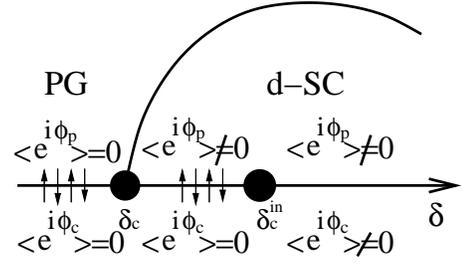}
\caption{\label{Fig. 1} Schematic phase diagram of underdoped high
$T_c$ cuprates at low energy in the plane of hole doping $\delta$
vs. temperature $T$. Only the pseudogap phase (PG) and $d-wave$
superconducting phase (d-SC) are partially shown.}
\end{figure}

Integrating over the Dirac spinon field and the holon field in
${\cal L}_{c}$, we obtain in the absence of electromagnetic field
$A_{\mu}$, \bqa &&{\cal L}_{c} = \frac{K}{2}|\partial_{\mu}\phi_c
+ 4a_{\mu}|^2 +
\frac{1}{2g_{eff}}(\partial\times{a})\frac{1}{\sqrt{-\partial^2}}(\partial\times{a})
, \eqa where $g_s = \frac{8}{N_s}$\cite{DonKim_CD}, $g_b =
\frac{8}{N_b}$\cite{Pisarski_CD} and $g_{eff} =
\frac{g_bg_s}{g_b+g_s}$. Here $N_s$ is the flavor number of the
Dirac fermions and $N_b$, the flavor number of the holon
quasiparticles. At the critical hole concentration $\delta_c$
where the superconducting phase transition occurs, the internal
gauge charge $CDT$ may not necessarily occur, as schematically
shown in Fig. 2\cite{Fradkin_CD,Senthil_CD,Sudbo_CD,NaLee_CD}. It
may be possible to have the internal charge $CDT$ at a critical
value $\delta_{c}^{in}$ of hole doping above $\delta_{c}$ where a
critical value $K_c$ of the chargeon pair phase stiffness is
reached to cause the condensation of chargeon pair bosons, i.e.,
$<e^{i\phi_{c}}> \not= 0$ and the suppression of the internal
gauge fluctuations $a_{\mu}$. On the other hand, below the
critical phase stiffness $K_c$ (the phase boundary line in Fig. 2)
the internal charge confinement phase will result
in\cite{Anomalous_gradient}.
\begin{figure}
\includegraphics[width=6cm]{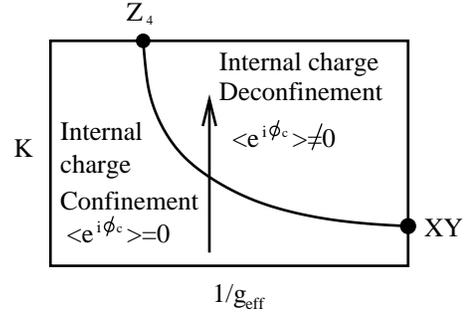}
\caption{\label{Fig. 2}Schematic phase diagram of the $(2+1)D$
Abelian Higgs model with multiple charge $4\tilde{e}$ in the plane
of the phase stiffness $K$ of the chargeon pair vs. the effective
coupling strength $g_{eff}$ shown in Eq. (4).}
\end{figure}

In passing, we briefly discuss the coexistence between the $AFM$
and $dSC$\cite{Kim2_CD} for completeness. In the underdoped region
of $\delta_{c} < \delta < \delta_{c}^{in}$ the $dSC$ coexists with
the confinement phase of the internal charge $\tilde{e}$ as is
schematically displayed in Fig. 2. The chargeon pair becomes
disordered, i.e., $<e^{i\phi_c}>=0$ below the critical phase
stiffness $K_c$ in this low hole doping region. This leads to
$AFM$ of the nodal fermions, as will be discussed here.
Integrating over $\phi_c$ in the disordered phase of
$<e^{i\phi_{c}}> = 0$, we obtain from Eq. (3), \bqa &&{\cal L}_{c}
= \bar\psi_l\gamma_\mu(\partial_\mu - ia_{\mu})\psi_l +
|(\partial_\mu - ia_{\mu} - iA_{\mu})\psi_b|^2 \nn && +
\frac{1}{2g}|4\partial\times{a} + 2\partial\times{A}|^2 \eqa with
the coupling strength $g \sim K_{c} - K$\cite{Kim2_CD}. Neglecting
coupling between the holon field and the gauge field $a_{\mu}$ for
the time being, the Dirac fermion field is known to be massive for
$N_{s} < N_{c} \approx 3.24$ in the case of non-compact U(1) gauge
field\cite{DonKim_CD,QED_CD,HerbutAF_CD,Tesanovic,Kim2_CD} where
$N_{s}$ is the flavor number of the Dirac fermion and $N_{c}$, the
critical flavor number for chiral symmetry breaking. The massive
Dirac fermion leads to $AFM$\cite{DonKim_CD,HerbutAF_CD,Kim2_CD}.
The magnetization $m$ is proportional to the effective coupling
strength $g$\cite{DonKim_CD,HerbutAF_CD}. Thus it is given by $m
\sim ({K}_{c} - K) \sim (\delta_{c}^{in2} -
\delta^2)$\cite{Kim2_CD}. The chargeon pair sector and the Cooper
pair sector are decoupled as shown in the effective Lagrangian Eq.
(3). This implies that the $AFM$ of the nodal fermions coexists
with the $dSC$ by having no direct coupling (interplay) between
the two\cite{Kim2_CD}. Now taking into account the contribution of
the holon field coupled to the internal gauge field, the critical
flavor number $N_{c} \approx 3.24$ of the chiral symmetry breaking
is reduced to $N_{c}' \approx 2.24$\cite{Flavor_number}. The
flavor number $N_s = 2$ of present interest is close to the
critical flavor number. Admitting the compactness of the internal
U(1) gauge field, confinement of the internal U(1) gauge charge is
not clearly understood yet. Recently, Hermele et. al claimed that
critical fluctuations of massless particles can result in
deconfinement of the internal charge in the limit of large flavor
numbers\cite{Hermele}. In our case the flavor number of
quasiparticles is small and thus the deconfinement is not expected
to occur from the massless quasiparticle excitations. Instead
particle-hole bound states are expected to occur, causing a
dynamically generated mass to both the spinons and holons. To
determine the critical flavor number in the presence of instantons
remains as an open issue.

Using the unitary gauge $4\tilde{a}_{\mu} = \partial_{\mu}\phi_{c}
+ 4a_{\mu} + 2A_{\mu}$ in Eq. (3), we obtain for ${\cal L}_{c}$
\bqa &&{\cal L}_{c} = \bar{\psi}_{l}\gamma_{\mu}(\partial_{\mu} +
i\frac{1}{4}\partial_{\mu}\phi_{c} - i\tilde{a}_{\mu} +
i\frac{1}{2}A_{\mu})\psi_{l} \nn && + |(\partial_{\mu} +
i\frac{1}{4}\partial_{\mu}\phi_{c} + i\tilde{a}_{\mu} -
i\frac{1}{2}A_{\mu})\psi_{b}|^2 + \frac{K}{2}|4\tilde{a}_{\mu}|^2.
\eqa In the large limit of stiffness parameter, $K \rightarrow
\infty$, the gauge fluctuations have $Z_4$
symmetry\cite{Fradkin_CD,Senthil_CD,Sudbo_CD,NaLee_CD}. Owing to
the internal gauge charge $4\tilde{e}$ (see the last term in Eq.
(6)) the effective field theory of the quasiparticles is involved
with the $Z_4$ gauge theory in association with the Dirac spinon
and the holon. In the deconfinement phase the $Z_4$ gauge
fluctuations are suppressed and coupling between the Dirac spinon
and the holon via the $Z_4$ gauge field becomes
negligible\cite{Senthil_CD}. Thus both the massless Dirac fermion
and the massless holon become free. The chiral symmetry breaking
is not expected to occur. Introducing the renormalized spinon
(quasi-spinon) field $\psi_{cf} =
e^{i\frac{1}{4}\phi_{c}}\psi_{l}$ and the renormalized holon
(quasi-holon) field $\psi_{cb} = e^{i\frac{1}{4}\phi_{c}}\psi_{b}$
in the above Lagrangian, an effective Lagrangian in the
deconfinement phase is obtained to be \bqa {\cal L}_{c} =
\bar{\psi}_{cf}\gamma_{\mu}(\partial_{\mu} +
i\frac{1}{2}A_{\mu})\psi_{cf} + |(\partial_{\mu} -
i\frac{1}{2}A_{\mu})\psi_{cb}|^2  . \eqa As can be seen from this
Lagrangian, the quasi-spinon carries the fractional  charge $-e/2$
and the quasi-holon, the fractional charge $+e/2$. Thus it is
possible that in the deconfinement phase charge fractionalization
may occur. This deconfinement phase corresponds to a
fractionalized metallic state of massless quasiparticles. The
observed antiferromagnetic to paramagnetic
transition\cite{Experiments} can be understood in the context of
the $CDT$. Various physical states involved with quantum phase
transitions are summarized in both Table 1 and Fig. 1.
\begin{table*}
\caption{Physical states in association with quantum phase
transitions}
\begin{tabular}{cccccccc}
\hline
  & Pseudogap  & Superconductivity  &  Superconductivity   \nn
  & ($\delta<\delta_{c}$) & ($\delta_{c} < \delta < \delta_{c}^{in}$) & ($\delta_{c}^{in}<\delta$) \nn
  \hline
  Phase &  $<e^{i\phi_p}> = 0$ & $<e^{i\phi_p}> \not= 0$  &  $<e^{i\phi_p}> \not= 0$  \nn
  & $<e^{i\phi_c}> = 0$  & $<e^{i\phi_c}> = 0$   &  $<e^{i\phi_c}> \not= 0$  \nn
  & $<\bar{\psi}_{l}\psi_{l}> \not= 0$ & $<\bar{\psi}_{l}\psi_{l}> \not= 0$  & $<\bar{\psi}_{l}\psi_{l}> = 0$ \nn
   Physical state & Mott insulator   & Superconductor  & Superconductor  \nn
  &Confinement  & Confinement   &  Deconfinement  \nn
    &  Antiferromagnetism  &  Antiferromagnetism  &  Paramagnetism
    \nn \hline
\end{tabular}
\end{table*}

At the critical point $\delta_{c}^{in}$ Algebraic Fermi liquid is
expected to occur since instanton excitations become irrelevant
owing to the gauge charge $4\tilde{e}$ of the chargeon pair
field\cite{Deconfined_critical_point}. Integrating over the
chargeon pair field at the critical point in Eq. (3) and
incorporating the gauge shift of $a_{\mu} \rightarrow a_{\mu} -
\frac{1}{2}A_{\mu}$, we obtain $QED_3$ in terms of the massless
Dirac spinon and massless holon coupled to the non-compact
internal U(1) gauge field,
\begin{eqnarray}
&&{\cal L}_{c} = \bar\psi_l\gamma_\mu(\partial_\mu - ia_{\mu} +
i\frac{1}{2}A_{\mu})\psi_l + |(\partial_\mu - ia_{\mu} -
i\frac{1}{2}A_{\mu})\psi_b|^2 \nn && +
\frac{N_a}{16}(\partial\times{a})\frac{1}{\sqrt{-\partial^2}}(\partial\times{a}),
\end{eqnarray}
where $N_a$ is the flavor number of the chargeon pair field. Here
$N_a = 1$. Even if Eq. (8) is similar to Eq. (5), the internal
U(1) gauge field is non-compact in Eq. (8) while it is compact in
Eq. (5). The chiral symmetry breaking is not expected to occur
since the coupling strength is above the critical value, i.e.,
$N_a + N_s + N_b > N_{c}$. This result is consistent with the fact
that the mass gap of the Dirac fermion or the magnetization
vanishes at the critical hole doping $\delta_{c}^{in}$ ($m \sim
\delta_{c}^{in2} - \delta^2$). As well known, there exist no well
defined quasiparticles in this $QED_3$ owing to massless gauge
fluctuations\cite{DonKim_CD}. But in the deconfinement phase the
renormalized spinon and holon are expected to be well defined
quasiparticles because the $Z_4$ gauge fluctuations are massive
and suppressed. This seems to be the main difference between
critical metallic state at the critical point and fractionalized
metallic state in the deconfinement phase. Our non-compact $QED_3$
emerging at the critical point in the superconducting state may
explain non-Fermi liquid behavior\cite{Tesanovic_AFL} in a
completely different context from the previous
studies\cite{Chubukov,MFL,CC_DDW}. This non-Fermi liquid behavior
is not related with the superconductivity but associated with the
$CDT$.

We showed that there exists no interplay between the
antiferromagnetism and superconductivity despite the coexistence.
This results in the fact that the superconducting transition falls
into the XY universality class in agreement with
experiments\cite{TypeII_CD}. In addition, we found that the
antiferro- to para- magnetic transition in the superconducting
state\cite{Experiments} can be resolved by the confinement to
deconfinement transition of the internal gauge charge. Further, we
argued that non-Fermi liquid behaviors in high $T_c$ cuprates may
be described by a critical field theory ($QED_3$) in association
with the new quantum critical point.

One of us (S.H.S.S) greatly acknowledges generous supports of
Hakjin program of Korean Ministry of Education (2003-2004) and the
BSRI program of Pohang University of Science and Technology
(2003-2004).


\begin{thebibliography}{9}
\bibitem{Renormalization} Performing a standard duality transformation of Eq. (2),
we obtain \bqa && {\cal L} = |(\partial_{\mu} -
ic_{c\mu})\Phi_{c}|^2 + V(|\Phi_{c}|) +
\frac{1}{2K}|\partial\times{c}_{c}|^2 \nn && -
i(\partial\times{c}_{c})_{\mu}(4a_{\mu} + 2A_{\mu}) \nn && +
\bar\psi_l\gamma_\mu(\partial_\mu - ia_{\mu})\psi_l +
|(\partial_\mu - ia_{\mu} - iA_{\mu})\psi_b|^2 \nn && +
|(\partial_{\mu} - ic_{p\mu})\Phi_{p}|^2 + V(|\Phi_{p}|) +
\frac{1}{2K}|\partial\times{c}_{p}|^2 +
i(\partial\times{c}_{p})_{\mu}2A_{\mu} \nn && +
\frac{1}{K_{cp}}(\partial\times{c}_{c})\cdot(\partial\times{c}_{p})
. \eqa Here $\Phi_{c(p)}$ is a chargeon (Cooper) pair vortex
field, and $c_{c(p)\mu}$, a chargeon (Cooper) pair vortex gauge
field mediating interactions between the vortices.
$V(|\Phi_{c(p)}|)$ is a vortex potential including vortex mass and
self-interaction terms. $K = \frac{K_{sp}K_{bp}}{K_{sp} + K_{bp}}$
is a renormalized phase stiffness resulting from the coupling
between the chargeon pair and Cooper pair fields, i.e.,
$\kappa_{cp}(\partial_\mu\phi_c + 4a_\mu +
2A_\mu)(\partial_\mu\phi_p - 2A_\mu)$. $K_{cp} =
\frac{K_{sp}K_{bp}}{K_{sp} - K_{bp}}$ is a residual coupling
between the two vortex gauge fields. The only remaining coupling
$\frac{1}{K_{cp}}(\partial\times{c}_{c})\cdot(\partial\times{c}_{p})$
between the Cooper pair sector and chargeon pair sector is
irrelevant in the renormalization group sense. The scaling
dimension of the coupling strength is $\left[ \frac{1}{K_{ac}}
\right] = -1$ and thus the strength of the coupling diminishes as
the cut-off energy gets smaller in the renormalization group
procedure. As a result the Cooper pair sector and chargeon pair
sector are completely decoupled.
\bibitem{Flavor_number} Owing to the massless holon excitations the
kinetic energy of the internal gauge field $a_{\mu}$ becomes
$\frac{N_s +
N_b}{16}(\partial\times{a})\frac{1}{\sqrt{-\partial^2}}(\partial\times{a})$
in the $1/N$ expansion. We note that $N_s$ in the absence of the
holon field is replaced by $N_s + N_b$ in the presence of that. In
our case the flavor number of the holon quasiparticle is $N_b =
1$. Thus the critical flavor number of the Dirac fermion is
obtained to be $N'_{c} = N_{c} - N_{b} \approx 2.24$.
\end{thebibliography}
\end{document}